# On the Transition from Gas-like to Liquid-like Behaviour in Supercritical $N_2$

J.E. Proctor[1*], C.G. Pruteanu[2*], I. Morrison[1], I.F. Crowe[3] and J.S. Loveday[4]

[1]Materials and Physics Research Group, University of Salford, Manchester M5 4WT, UK

[2]Department of Physics and Astronomy, University College London, Gower Street, London WC1E 6BT, UK

[3]Photon Science Institute and School of Electrical and Electronic Engineering, University of Manchester, Oxford Road, Manchester M13 9PL

[4]SUPA, School of Physics and Astronomy and Centre for Science at Extreme Conditions, The University of Edinburgh, Edinburgh, EH9 3JZ, UK

*These authors contributed equally to the work

**Abstract**

We have studied in detail the transition from gas-like to rigid liquid-like behaviour in supercritical $N_2$ at 300 K (2.4 $T_C$). Our study combines neutron diffraction and Raman spectroscopy with ab-initio molecular dynamics simulations. We observe a narrow transition from gas-like to rigid liquid-like behaviour at ca. 150 MPa, which we associate with the Frenkel line. Our findings allow us to reliably characterize the Frenkel line using both diffraction and spectroscopy methods, backed up by simulation, for the same substance. We clearly lay out what parameters change, and what parameters do not change, when the Frenkel line is crossed.

The traditional textbook view of liquids is that matter does not exist in a liquid, or liquid-like state, above the critical temperature $T_C$. However, in recent years two separate ideas have been proposed that both contradict this viewpoint. On the one hand, rapid fluctuations in certain thermodynamic properties roughly following the critical isochore close to the critical point, the Widom lines, have been proposed to extend much further beyond $T_C$ than previously believed on the basis of the fundamental equation of state (EOS) (1) and to mark a transition to a liquid-like state (1-3). Another transition emanating from the critical point has been proposed very recently (4). These transitions are essentially thermodynamic continuations of the vapour pressure curve.

On the other hand, it is argued that our theoretical understanding of liquids relies excessively on comparison to gases. For instance, a variety of cubic $PVT$ EOS are widely used to model fluid properties – Patel-Teja, Redlich-Kwong, Redlich-Kwong-Soave, Peng-Robinson, etc. All these EOS are empirical improvements to the Van der Waals EOS which was, in turn, an adaptation of the ideal gas EOS by introducing terms to account for the finite volume of the gas particles and attraction between gas particles. These equations are fundamentally different from the EOS commonly utilized to describe the properties of solids.

Adoption of the alternate approach of understanding the liquid state by comparison to solids at $P,T$ conditions close to crystallization has led to the proposal that the liquid state can be divided into regions where the liquid behaviour is gas-like (we can use the term non-rigid liquid) and where the liquid behaviour is solidlike (rigid liquid). The proposed narrow (but not first order) dividing line between these regions is called the Frenkel line (5). The Frenkel line is proposed to continue beyond $T_C$ so, in the Frenkel line vision whilst the non-rigid liquid state does not exist beyond $T_C$ the rigid liquid state can exist at arbitrarily high temperature if the density is kept high enough. The Frenkel line was originally proposed as a transition in certain dynamic properties, for instance speed of sound and heat capacity (5), rather than static properties. Later, however, it was pointed out (6) that one cannot separate consideration of static and dynamic properties in this manner. The two are intrinsically linked for all matter, for instance through the energy equation (equation 1) (7) for fluids linking the internal energy $U$ to the pair distribution function $g(r)$ via the pair potential $V(r)$. Here, $U_K$ refers to the kinetic component of $U$ ($3k_BT/2$ for a monatomic fluid) and the integral gives the potential energy component.

$$U = U_K + 2\pi\rho \int_0^\infty V(r)g(r)r^2dr \quad (1)$$

Thus, there is now a consensus that a liquid, or liquid-like state, can exist beyond $T_C$ but fundamental disagreement about the density required, and mechanism, for a transition into a liquid-like state. The Frenkel line is expected at pressures an order of magnitude higher than the Widom lines. The Widom lines unquestionably exist, yet their utility in defining a liquid-like state has been questioned due to the rapid divergence of different Widom lines and the smearing out of these transitions as $P,T$ is increased significantly beyond the critical point(8,9). In contrast, experimental studies claiming to observe the Frenkel line are scarce(6,10,11) and have been disputed(12,13). Bryk et al. claimed to "critically undermine" the concept of the Frenkel line (13).

Experimental work to date has been hampered by the adoption of erratic $P,T$ paths, the well-documented $Q_{max}$-cutoff problem in X-ray diffraction (14) and difficulties in achieving the required accuracy in pressure control and measurement inside the diamond anvil cell (DAC), especially at high temperature (10). These problems have fuelled the controversy. This matter is important not just due to the industrial applications of supercritical fluids, but because the $P, T$ paths in the outer layers of the major and outer planets are expected to pass right through this part of the phase diagram.

Here, we conduct a study of the transition from gas-like to rigid liquid-like behaviour across the Frenkel line in supercritical $N_2$ in which the transition is characterized in unprecedented detail by the combination of neutron diffraction (to circumvent the $Q_{max}$-cutoff problem) and Raman spectroscopy on the same sample, supported by ab-initio molecular dynamics (MD). Empirical potential structure refinement (EPSR) was performed to reliably extract pair distribution functions and coordination numbers from the measured total scattering data. This allows us to clarify what changes, and what does not change, when the Frenkel line is crossed. Our results can provide a practical experimental definition of the Frenkel line to guide future studies.

Experimental and simulation methods are outlined in the supporting information. Figure 1 (a) shows a representative selection of the raw $S(Q)$ data obtained, including the lowest (25 MPa) and highest (300 MPa). The general form of the $S(Q)$ data obtained remains similar throughout. However, the first peak moves outwards, increases in amplitude, and sharpens with increasing pressure. A shoulder on the right side of the first peak moves to higher Q and becomes a new peak. Because it involves the largest relative density change the pressure increment from 25 to 75 MPa produces the largest changes in S(Q). Figure 1 (b) shows the molecular (excluding intra-molecular features) $g(r)$ corresponding to these $S(Q)$ data, obtained using EPSR. Note that these $g(r)$ data are not simply direct Fourier transforms of the $S(Q)$ data. They are produced from the arrangement of molecules in the EPSR "box" that produces an $S(Q)$ most closely fitting the diffraction data. In the $g(r)$ data we also observe the most significant changes between 25 MPa and 75 MPa as expected. The main changes are the development of the first neighbour peak at around 4 Å followed at high pressures by the second and third neighbour peaks as might be expected for a fluid that is becoming more liquid-like. There is a sharp peak at $\sim$2.5 Å which appears only in the 25 MPa data. This peak is present only in this data set and makes no physical sense. It also appears in the raw Fourier transform of the $S(Q)$. For these reasons we attribute this peak to an artefact due to the poor signal-to-noise ratio in the $S(Q)$ data in this low-density sample. Figure 1 (c) shows the variation of the co-ordination number as a function of pressure, calculated using the position of the minimum in $g(r)$ between the first and second neighbour peaks as a cut-off radius. As can be seen, the co-ordination rises quickly from 9 to $\approx$11.5 at 175 MPa and thereafter remains constant. This is consistent with the idea that above this pressure nitrogen has the co-ordination of a close packed dense liquid. The co-ordination number at 25 MPa includes the effect of the spurious 2.5 Å peak (see above) and may therefore overestimate the co-ordination at this pressure. To assess the magnitude of this overestimate we carried out integrations of the co-ordination with the peak removed and found that the co-ordination reduced by only 0.5 molecules. We have chosen to use all our data for figure 1 (c), but note that the effect of the spurious peak is small and does not affect our conclusions. We have also examined the variation of other parameters (position and width of the peaks) in the $g(r)$ data as a function of pressure.  Within the scatter in the data, these vary smoothly throughout (see supporting information).

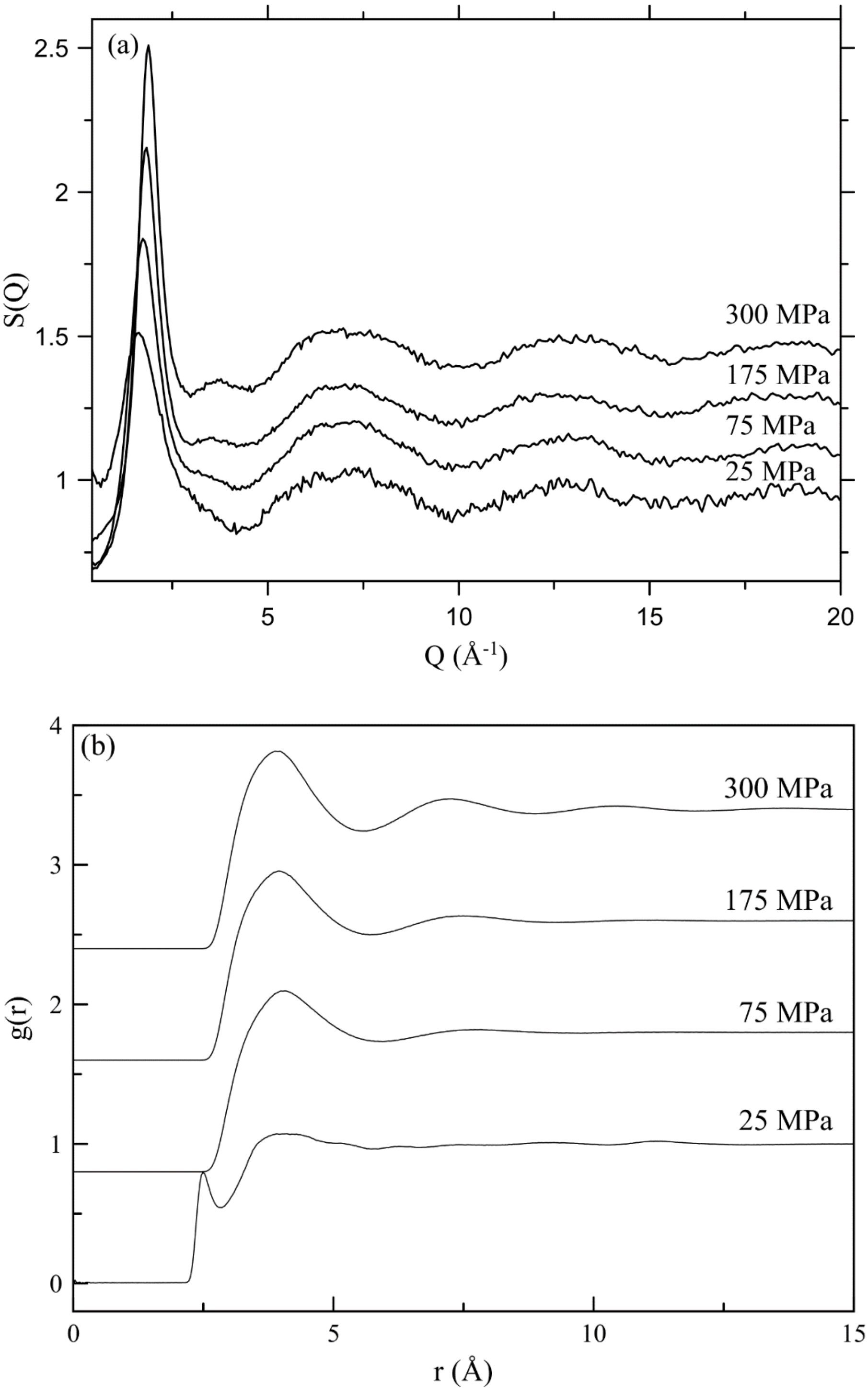

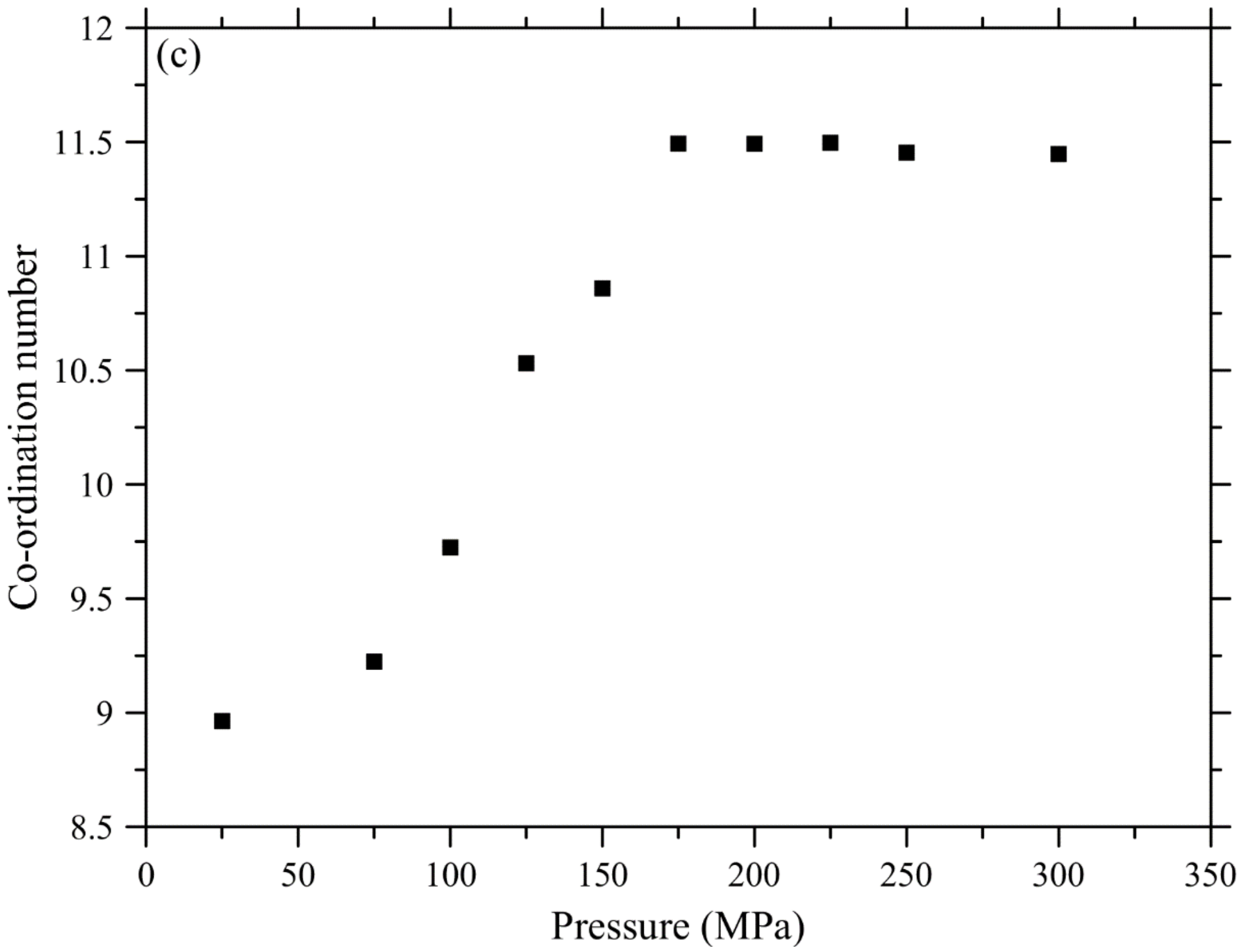

Figure 1. (a) Selected plots (offset by 0.2) of obtained diffraction data $S(Q)$. (b) Selected plots of $g(r)$ (offset by 0.8) extracted using empirical potential structure refinement (EPSR). (c) Plot of the co-ordination number for the first shell as a function of pressure, obtained using EPSR by numerical integration of the first shell in the $g(r)$ function.

The EPSR process also determines the (intra-molecular) N-N bond length at each pressure. We would expect this to decrease at the higher pressures studied. This expected trend is observed, although it is barely visible due to the scatter in the data (see supporting information). We note also that the data cannot be quantitatively reconciled with the accepted gas phase bond length $a_0 = 1.09768$ Å (15). Throughout, the bond length given by EPSR is ca. 1% too long.

We also collected information on the N-N bond length $a$ using Raman scattering. The frequency of the intense Raman-active vibron in $N_2$ at ca. 2330 cm$^{-1}$ can be linked to the bond length via the dimensionless mode Grüneisen parameter $\gamma$ (equation 2). Whilst the shift in bond length and Raman frequency relative to their absolute values is extremely small, the change in Raman frequency is easily resolvable, allowing the collection of data on the bond length with much higher resolution (figure 2) than is possible using diffraction measurements and EPSR.

$$\frac{\omega(P)}{\omega_0} = \left[\frac{a(P)}{a_0}\right]^{-\gamma} \qquad (2)$$

The mode Grüneisen parameter $\gamma$ is a measure of the influence of anharmonic effects on the vibrational frequency of this specific vibrational mode and can be calculated from the pair potential for the N – N bond. The Lennard-Jones (6-12) potential is exactly correct only for Van der Waals bonding, but is a good approximation which is used successfully for covalent bonding (16)(17). The use of any potential of the 6-12 form results in $\gamma = 3.5$ exactly so we have used this value in our analysis. In figure 2 (a) we show a plot of Raman frequency versus pressure at 300 K, combining our data with that of May et al. (18) and selected data from Devendorf and Ben-Amotz (15). Data from other studies (19-21) are consistent with these findings. Figure 2 (b) shows the obtained values of N – N bond length $a$ as a function of pressure. The gas phase bond length has been obtained from ref. (15).

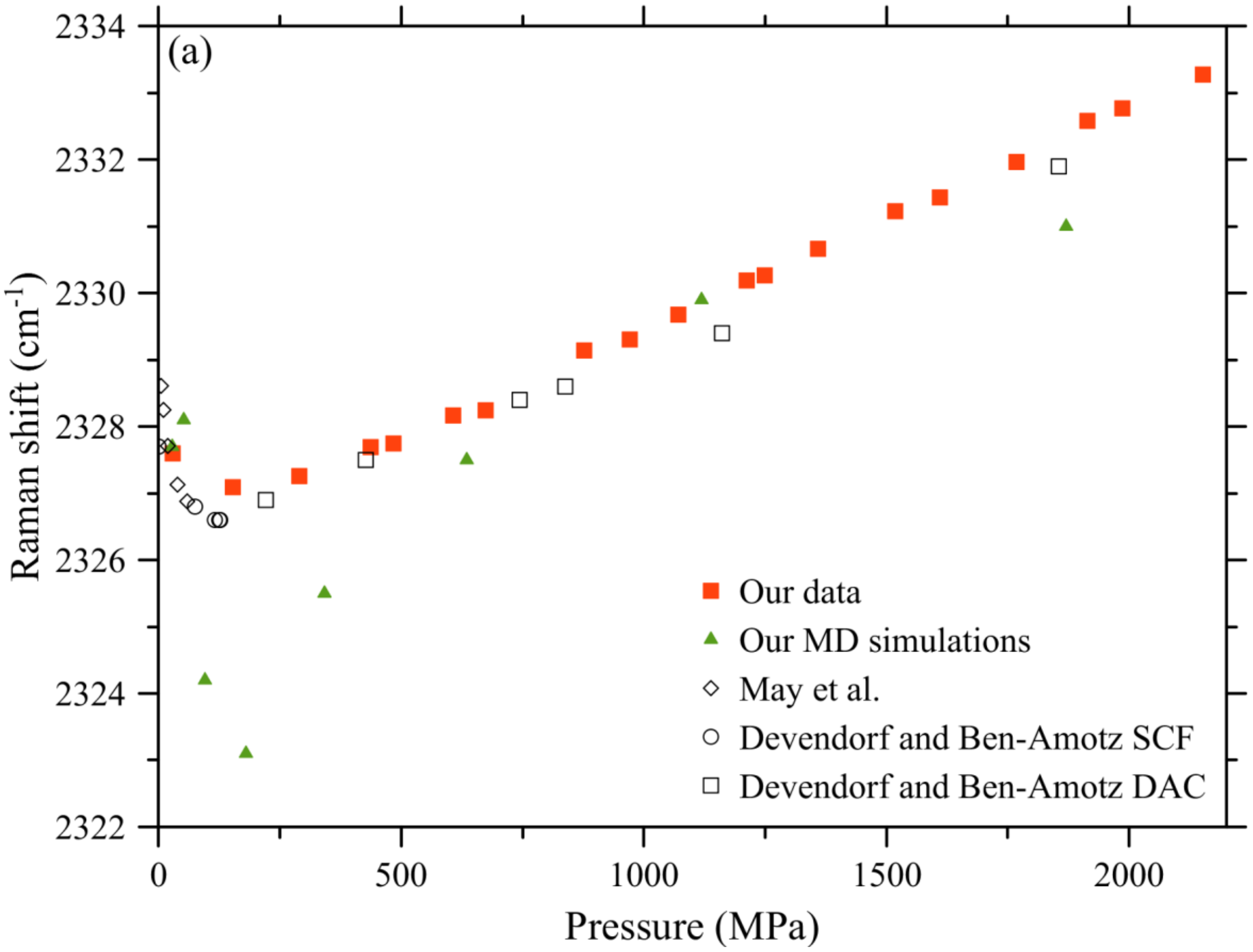

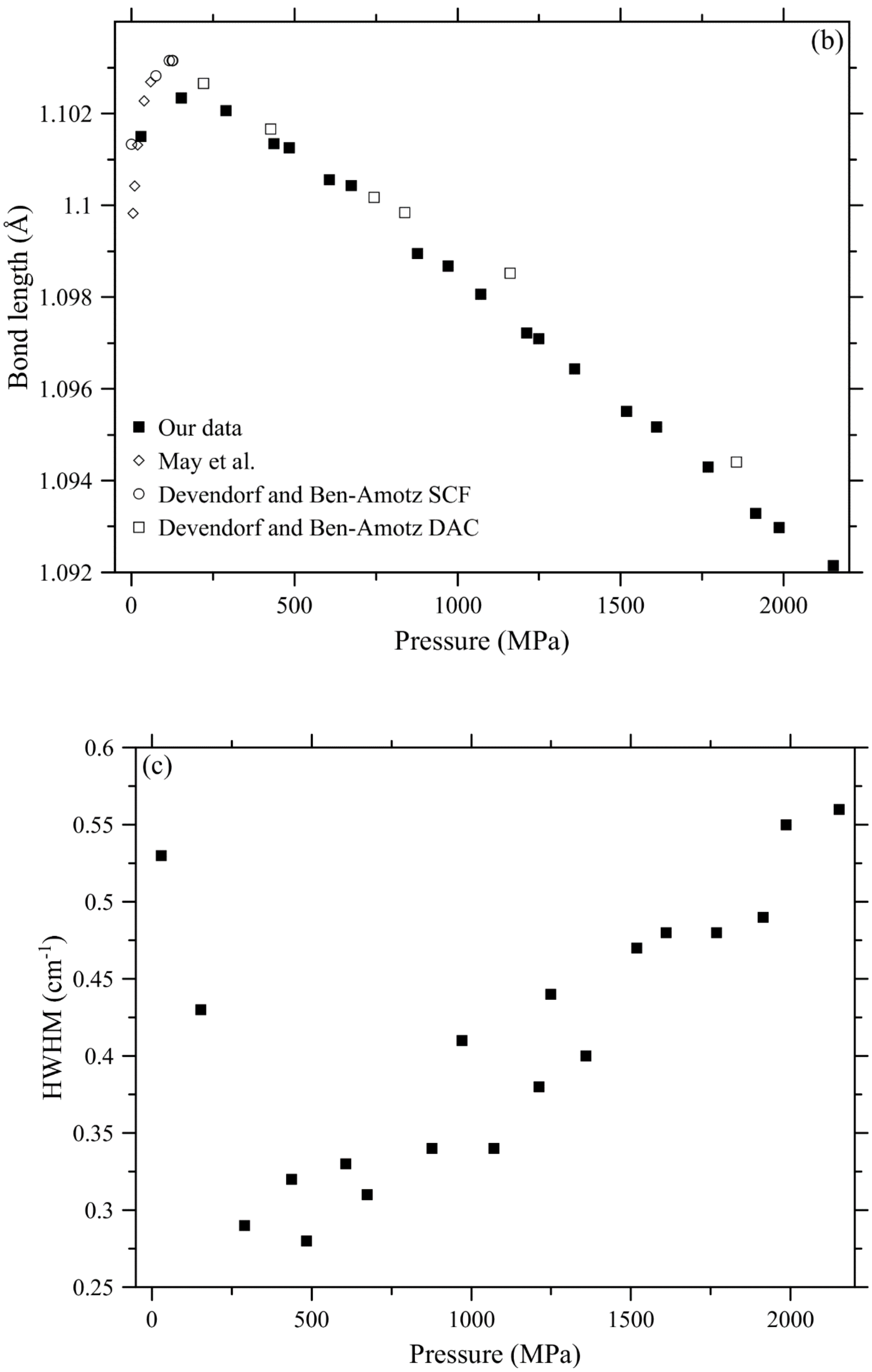

Figure 2. (a) Raman peak position as a function of pressure, plotted from present data and MD simulations alongside previous studies in the diamond anvil cell (DAC) and supercritical fluid cell (SCF) (refs. (15,18) and refs. therein). (b) N – N bond length $a$ as a function of pressure, calculated from Raman peak position using equation (2). (c) Raman peak HWHM as a function of pressure.

The Raman peak width (half width half maximum, HWHM) (figure 2 (c)) also displays qualitatively different pressure dependencies at low density and high density. Unfortunately the peak width is not given in ref. (15) but May et al. (18) observed a decreasing linewidth upon pressure increase up to the highest pressure reached in their study (59 MPa). We also observe a decreasing linewidth upon pressure increase at the lowest pressures studied, followed by an increase in linewidth as pressure is increased beyond ca. 250 MPa. We cannot quantitatively compare our results to May et al. since the narrowest linewidths observed are close to the limit set by the resolution of the spectrometers used in the different studies.

We can state little about the intensity variation as a function of pressure since this is not given in refs. (15,18). In our own data the intensity (normalized to account for the density change) exhibits a steady increase upon pressure increase throughout, albeit with some noise in the data (see supporting information).

Our experimentally determined Raman peak position is in good agreement with the results of our ab-initio MD simulations (figure 2(a)). Whilst the MD simulations slightly overestimate the effect on Raman frequency of the changes in density the trends observed agree nicely with experiment. In addition, we used ab-initio MD to examine if any kind of orientational order existed in the sample. This was not the case at any of the densities studied.

Analysis of our combined data demonstrates that there exists a narrow transition between gas-like and rigid liquid-like behaviour at ca. 150 MPa, which we associate with the Frenkel line, and which corresponds to where the Frenkel line is expected to lie in $N_2$ at this temperature on the basis of available theory and heat capacity measurements. The position of the Frenkel line can be estimated in the supercritical region from a consideration of the heat capacity at constant volume $C_V$. The definition of the Frenkel line in terms of shear wave propagation is that the line is crossed when $C_V = 2k_B$ (5), following (in this case) subtraction of the contribution from rotational motion ($k_B$). According to this definition, the Frenkel line is crossed at 150.0 MPa at 300 K. The Frenkel line according to this criterion is plotted in figure 3 down to 160 K, beneath which the heat capacity criterion can no longer be used due to proximity to the peak in heat capacity due to the Widom line for $C_V$, which is very broad by this temperature. The Frenkel line position at 300 K has also been estimated by rescaling of the Lennard-Jones parameters used to estimate the Frenkel line position for Ar (22). This estimate puts the Frenkel line at 300 K at 140 MPa. Thus both estimates that can be made from the literature on the Frenkel line position at 300 K are in excellent agreement with our data and simulations. At this pressure the co-ordination number plateaus out at the expected maximum value of just under 12 ($N_2$ crystallizes into the hexagonal close-packed structure with a co-ordination number of 12 at 300 K). The behaviour of the Raman peak position demonstrates that contraction of the intra-molecular bond is taking place above 150 MPa to achieve further compression. The linewidth undergoes collisional narrowing up to ca. 150 MPa as also characterized in the previous study to 59 MPa (18), then increases upon further pressure increase. This can be explained by the fluid, whilst being unable to support static shear stress, being able to support shear (nonhydrostatic) stress for a limited time due to the propagation of shear waves that is possible on the rigid liquid side of the Frenkel line (5,11).

Thus, up to 150 MPa behaviour is gas-like, in the sense that compression of the sample is achieved by reduction of the inter-molecular distance and increase in the co-ordination number. In this regime, we obtain the slightly counter-intuitive result that the bond length increases upon pressure increase. This is because the increase in density increases the influence of attractive Van der Waals forces between molecules, thus loosening the intra-molecular bond. Beyond 150 MPa the behaviour is rigid liquid-like, in the sense that the co-ordination number stays constant and much of the compression is achieved by direct compression of the N – N intra-molecular bond.

In conclusion, we have performed a careful and rigorous study of the manner in which supercritical fluid $N_2$ transforms from gas-like behaviour to rigid liquid-like behaviour at 300 K, by crossing the Frenkel line. The Frenkel line is crossed at 150 MPa at 300 K. Our neutron diffraction and Raman data, supported by MD simulations, are all in agreement on this point. These findings allow us to clearly and reliably quantify, for the first time, what changes and what does not change in the fluid diffraction data when the Frenkel line is crossed. These findings can guide future studies. As far as the diffraction data are concerned, the only parameter that could be reliably determined which exhibits a non-monotonic change when the Frenkel line is crossed is the co-ordination number. This stops increasing upon pressure increase and stays constant, with a value just below that for the solid phase into which the sample will crystallize upon further pressure increase.

Our findings contrast with those from previous studies utilizing X-ray diffraction to observe the Frenkel line(6,11), which observed a variety of other changes when the Frenkel line is claimed to be crossed, including in the position of the first peak in $g(r)$ (11). We suggest that these changes may be a result of the irregularity of the $P, T$ path followed in some cases in the experiments, and of the potential for unavoidable systematic errors in the background subtraction and Fourier transform procedures utilized to extract the peak parameters. The $Q_{max}$-cutoff problem, in particular, is well understood (14) and is an unavoidable consequence of the X-ray wavelength being similar to the interatomic spacing in order to obtain a diffraction pattern.

We therefore propose neutron diffraction, avoiding the $Q_{max}$-cutoff problem, as the reliable method to evaluate the location and character of the Frenkel line. If X-ray diffraction must be employed, we propose that the data should be analysed using EPSR in a manner similar to that employed in the present study or deploy similar methods (such as Reverse Monte Carlo) rather than by a direct Fourier transform of the diffraction data, to minimize the potential for systematic analysis errors. In addition, analysis should focus primarily on the co-ordination number since this quantity is calculated from the first peak in $g(r)$ so is less affected by the $Q_{max}$-cutoff problem. There is still the potential for a systematic error due to the choice of where to define the minimum between the first and second peaks in $g(r)$. We therefore agree with the proposal that the relative, rather than absolute, co-ordination number should be considered as the reliably determined quantity (23). In both our work and ref. (6) the co-ordination number plateaus out at a fixed value where we expect the Frenkel line to be crossed.

Reliably evaluating the presence (or lack thereof) of the Frenkel line solely through a diffraction experiment is difficult. We therefore propose that, where possible, diffraction should be combined with optical spectroscopy. In the Raman results reported here, and in similar experiments we have conducted previously on supercritical methane (10) and subcritical ethane (24), we observe behaviour which is fundamentally different in the gas-like and rigid liquid-like regions. The changes are evident in the raw data, are in agreement with our MD simulations, and their observation is not dependent on the validity of any analysis procedure that we may wish to employ.

The lack of orientational order existing in our MD simulations is interesting, but consistent with the limited orientational order in the corresponding solid phase of $N_2$ produced upon crystallization at 300 K. Bearing in mind that the drastic change in Raman frequency associated with the Frenkel line is reproduced by the MD simulations, we can conclude that orientational order does not necessarily need to exist on the rigid liquid-like side of the Frenkel line.

Finally, we would like to examine the location of the Widom lines for $N_2$ in relation to our data (figure 3). The Widom lines can be plotted out using the fundamental EOS (25) and (in common with Widom lines in other samples) stay close to the critical isochore until they terminate. As noted in the introduction, the Widom lines diverge from each other before terminating. The exact point at which a Widom line terminates is not clearly defined – the extremal value in the property simply gets more and more smeared out the further you go from the critical point. Our decisions of where to terminate the representative Widom lines in figure 3 are outlined in the supporting information. Other Widom lines that we examined ($C_V$, $C_P$, sound speed) all terminated at lower temperature than those shown in figure 3.

From the data summarized in figure 3 we may note a fundamental difference between the Frenkel line and the Widom lines. The former is crossed at far higher pressure and density (note the logarithmic pressure axis in figure 3). The Widom lines stay close to the critical isochore, at which there is enough empty space between adjacent particles in the fluid to squeeze in an additional particle. The Frenkel line, on the other hand, is crossed when the fluid has a density and co-ordination number close to that of a solid.

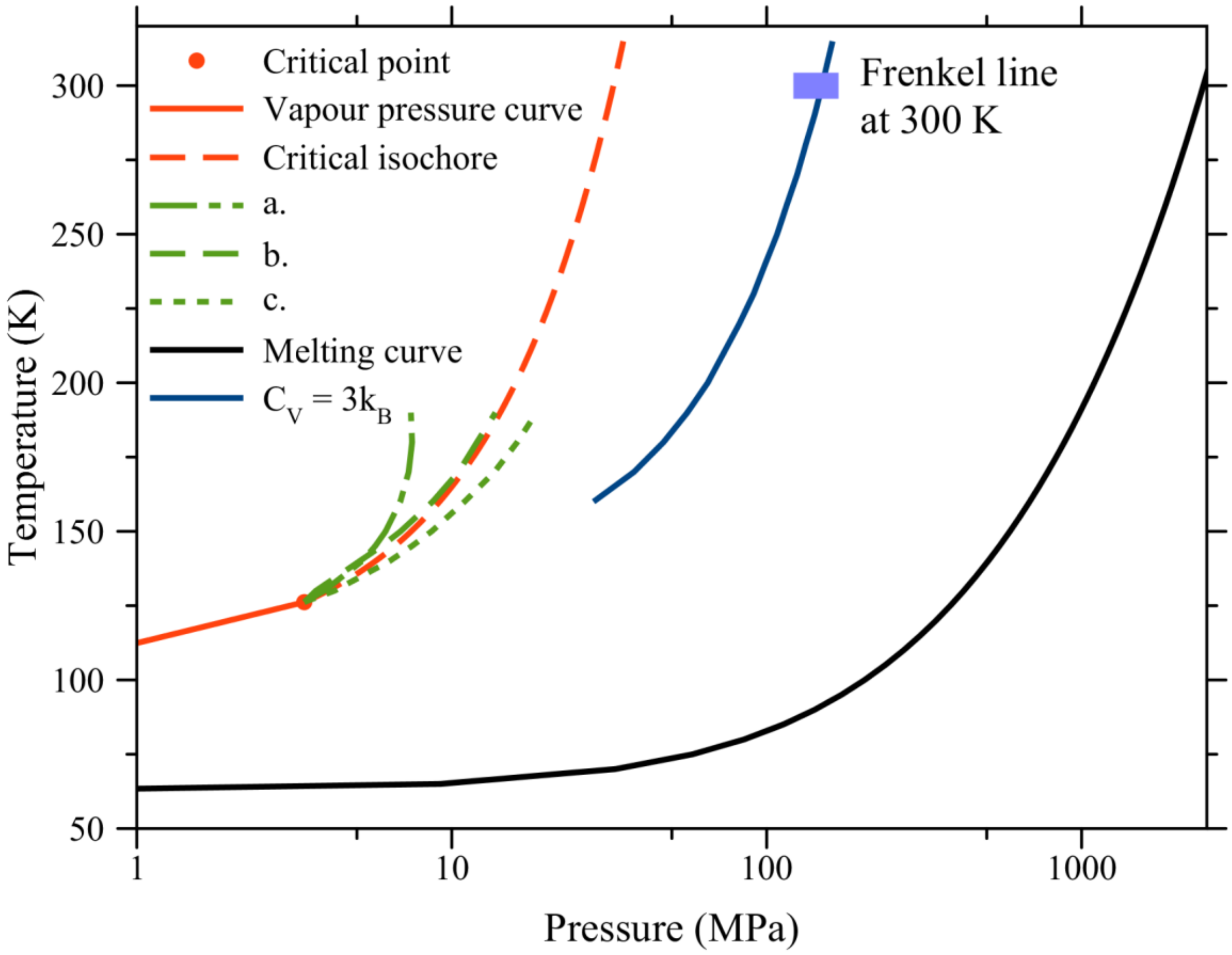

Figure 3. Phase diagram of fluid $N_2$. The green lines are the Widom lines extending furthest from the critical point: a. compressibility, b. thermal conductivity and c. viscosity.

It is clear from this that it is the Widom lines and related phenomena, rather than the Frenkel line, that dominate fluid behaviour in the region close to the critical point and result in large fluctuations in key fluid parameters in this region. They are therefore most important from the point of view of industrial applications of supercritical fluids which utilize this region of the phase diagram rather than the region at much higher density where the Frenkel line is crossed.

On the other hand, we have a fundamental theoretical objection to the argument that the Widom line(s) can mark a boundary between gas-like and liquid-like behaviour significantly above $T_C$. By definition, $T_C$ is defined by the relation between the available thermal energy and the depth $\varepsilon$ of the potential well due to the attractive potential between fluid particles at modest separations (equation 3) (26).

$$k_B T_C \approx \varepsilon \tag{3}$$

Thus, above $T_C$, the attraction between particles cannot cause condensation to a liquid-like state because sufficient thermal energy is always available to overcome the attractive potential and allow the particles to escape. The only way to force a sample into a liquid-like state significantly above $T_C$ is to increase density up to the point where there is nowhere for the particles to escape to. Our data on the co-ordination number (figure 1c) show that this condition is not met until the Frenkel line is crossed at ca. 150 MPa, at just over double the density at the critical isochore where the Widom lines lie. Below 150 MPa, all our data are consistent with the sample behaving in a gas-like manner in terms of both static and dynamic properties.

**Acknowledgements** We would like to acknowledge the provision of beamtime on the SANDALS instrument at the ISIS pulsed neutron source (beamtime RB1810145) and the assistance of Dr. Silvia Imberti at ISIS.

**References**

(1) G.G. Simeoni, T. Bryk, F.A. Gorelli, M. Krisch, G. Ruocco, M. Santoro and T. Scopigno, The Widom Line as the Crossover Between Liquid-like and Gas-like Behaviour in Supercritical Fluids, Nat. Phys. **2010**, *6*, 503-507.

(2) T. Bryk, F. Gorelli, G. Ruocco, M. Santoro and T. Scopigno, Collective excitations in soft-sphere fluids, Phys. Rev. E **2014**, *90*, 042301.

(3) F.A. Gorelli, T. Bryk, M. Krisch, G. Ruocco, M. Santoro and T. Scopigno, Dynamics and Thermodynamics beyond the Critical Point, Sci. Rep. **2013**, *3*, 1203.

(4) E.A. Ploetz and P.E. Smith, Gas or Liquid? The Supercritical Behavior of Pure Fluids, J. Phys. Chem. B **2019**, *123*, 6554-6563.

(5) V.V. Brazhkin, Yu.D. Fomin, A.G. Lyapin, V.N. Ryzhov and K. Trachenko, Two Liquid States of Matter: A Dynamic Line on a Phase Diagram, Phys. Rev. E **2012**, *85*, 031203.

(6) C. Prescher, Yu.D. Fomin, V.B. Prakapenka, J. Stefanski, K. Trachenko and V.V. Brazhkin, Experimental Evidence of the Frenkel Line in Supercritical Neon, Phys. Rev. B **2017**, *95*, 134114.

(7) J.P. Hansen and I.R. McDonald, Theory of Simple Liquids; Academic Press: Oxford, U.K.; 2013.

(8) Yu.D. Fomin, V.N. Ryzhov, E.N. Tsiok and V.V. Brazhkin, Thermodynamic Properties of Supercritical Carbon Dioxide: Widom and Frenkel Lines, Phys. Rev. E **2015**, *91*, 022111.

(9) A.R. Imre, U.K. Deiters, T. Kraska and I. Tiselj, The Pseudocritical Regions for Supercritical Water, Nucl. Eng. Design **2012**, *252*, 179.

(10) D. Smith, M.A. Hakeem, P. Parisaides, H.E. Maynard-Casely, D. Foster, D. Eden, D.J. Bull, A.R.L. Marshall, A.M. Adawi, R. Howie et al., Crossover Between Liquidlike and Gaslike Behavior in $CH_4$ at 400 K, Phys. Rev. E **2017**, *96*, 052113.

(11) D. Bolmatov, M. Zhernenkov, D. Zav'yalov, S.N. Tkachev, A. Cunsolo and Y.Q. Cai, The Frenkel Line: A Direct Experimental Evidence for the New Thermodynamic Boundary, Sci. Rep. **2015**, *5*, 15850.

(12) V.V. Brazhkin and J.E. Proctor, Comment on: "The Frenkel Line: a Direct Experimental Evidence for the New Thermodynamic Boundary" https://arxiv.org/abs/1608.06883 (2016).

(13) T. Bryk, F.A. Gorelli, I. Mryglod, G. Ruocco, M. Santoro and T. Scopigno, Behavior of Supercritical Fluids across the "Frenkel Line", J. Phys. Chem. Lett. **2017**, *8*, 4995-5001.

(14) J.H. Eggert, G. Weck, P. Loubeyre and M. Mezouar, Quantitative Structure Factor and Density Measurements of High-pressure Fluids in Diamond Anvil Cells by X-ray Diffraction: Argon and Water, Phys. Rev. B **2002**, *65*, 174105.

(15) G.S. Devendorf and D. Ben-Amotz, Vibrational Frequency Shifts of Fluid Nitrogen up to Ultrahigh Temperatures and Pressures, J. Phys. Chem. **1993**, *97*, 2307-2313.

(16) J.G. Powles and K.E. Gubbins, The Intermolecular Potential for Nitrogen, Chem. Phys. Lett. **1976**, *38*, 405-406.

(17) C.S. Murthy, K. Singer, M.L. Klein and I.R. McDonald, Pairwise Additive Effective Potentials for Nitrogen, Mol. Phys. **1980**, *41*, 1387-1399.

(18) A.D. May, J.C. Stryland and G. Varghese, Collisional Narrowing of the Vibrational Raman Band of Nitrogen and Carbon Monoxide, Can. J. Phys. **1970**, *48*, 2331-2335.

(19) C.H. Wang and R.B. Wright, Effect of Density on the Raman Scattering of Molecular Fluids. I. A Detailed Study of the Scattering Polarization, Intensity, Frequency Shift, and Spectral Shape in Gaseous $N_2$, J. Chem. Phys. **1973**, *59*, 1706-1712.

(20) R. Kroon, M. Baggen and A. Lagendijk, Vibrational Dephasing in Liquid Nitrogen at High Densities Studied with Timeresolved Stimulated Raman Gain Spectroscopy, J. Chem. Phys. **1989**, *91*, 74-78.

(21) M.I.M. Scheerboom, J.P.J. Michels and J.A. Schouten, High Pressure Study on the Raman Spectra of Fluid Nitrogen and Nitrogen in Helium, J. Chem. Phys. **1996**, *104*, 9388-9400.

(22) V.V. Brazhkin, A.G. Lyapin, V.N. Ryzhov, K. Trachenko, Yu.D. Fomin and E.N. Tsiok, Frenkel Line and Supercritical Technologies, Supercritical Fluids: Theory and Experiment **2014**, *9*, 40-50 (in Russian).

(23) G. Weck, J. Eggert, P. Loubeyre, N. Desbiens, E. Bourasseau, J.-B. Maillet, M. Mezouar and M. Hanfland, Phase Diagrams and Isotopic Effects of Normal and Deuterated Water Studied via X-ray Diffraction up to 4.5 GPa and 500 K, Phys. Rev. B **2009**, *80*, 180202(R).

(24) J.E. Proctor, M. Bailey, I. Morrison, M.A. Hakeem and I.F. Crowe, Observation of Liquid−Liquid Phase Transitions in Ethane at 300 K, J. Phys. Chem. B **2018**, *122*, 10172-10178.

(25) R. Span, E.W. Lemmon, R.T. Jacobsen, W. Wagner and A. Yokozeki, A Reference Equation of State for the Thermodynamic Properties of Nitrogen for Temperatures from 63.151 to 1000 K and Pressures to 2200 MPa, J. Phys. Chem. Ref. Data **2000**, *29*, 1361-1433.

(26) B.H. Flowers and E. Mendoza, Properties of Matter; Wiley-VCH: Chichester, U.K.; 1970.

(27) A.K. Soper, Computer Simulation as a Tool for the Interpretation of Total Scattering Data from Glasses and Liquids, Mol. Sim. **2012**, *38*, 1171-1185.

(28) S.J. Clark, M.D. Segall, C.J. Pickard, P.J. Hasnip, M.J. Probert, K. Refson and M.C. Payne, First Principles Methods using CASTEP, Z. Krystall. **2005**, *220*, 567-570.

(29) A. Tkatchenko and M. Scheffler, Accurate Molecular Van Der Waals Interactions from Ground-State Electron Density and Free-Atom Reference Data, Phys. Rev. Lett. **2009**, *102*, 73005.

Supporting Information for:

# On the Transition from Gaslike to Liquidlike Behaviour in Supercritical $N_2$

J.E. Proctor[1*], C.G. Pruteanu[2*], I. Morrison[1], I.F. Crowe[3] and J.S. Loveday[4]

[1]Materials and Physics Research Group, University of Salford, Manchester M5 4WT, UK

[2]Department of Physics and Astronomy, University College London, Gower Street, London WC1E 6BT, UK

[3]Photon Science Institute and School of Electrical and Electronic Engineering, University of Manchester, Oxford Road, Manchester M13 9PL

[4]SUPA, School of Physics and Astronomy and Centre for Science at Extreme Conditions, The University of Edinburgh, Edinburgh, EH9 3JZ, UK

*These authors contributed equally to the work

**Experimental and analysis methods**

Raman spectroscopy

A diamond anvil cell (DAC) was equipped with diamonds having 1 mm culets and a stainless steel gasket, indented then drilled using a custom-constructed spark eroder. Liquid $N_2$ was loaded into the sample chamber by immersing the entire DAC in liquid $N_2$ then turning the pressure screws using long Allen keys to close the DAC. Pressure was measured using the ruby photoluminescence method, resulting in a typical error of ±0.002 GPa. Pressure was measured both before and after the collection of the Raman spectra at each step to evaluate the total error in the pressure measurement. In all cases, error bars are too small to display. Data were collected on pressure decrease from 2.6 GPa at constant temperature. To calibrate the pressure measurement, a spectrum was collected of the same chip of ruby as used for the pressure measurement at ambient conditions.

Spectra of the intense Raman-active vibron at ca. 2330 $cm^{-1}$ were collected on a Renishaw InVia Raman spectrometer (633 nm laser excitation, 2800 lines per mm grating). The total instrumental linewidth was confirmed from the Raman spectrum of the unstressed diamond to be, at most, 1.3 $cm^{-1}$ HWHM. Each spectrum was fitted using a single Lorentzian peak in Magicplot Pro following subtraction of a linear background. The calibration of the Raman measurements was checked using the intense Raman-active vibration of diamond at 1332 $cm^{-1}$ at ambient conditions. Our data are in good agreement with those collected in previous studies in the pressure ranges in which they overlap, confirming the accuracy of this calibration.

## Neutron diffraction

Time-of-flight neutron diffraction data were collected on the SANDALS instrument at the ISIS pulsed neutron source. The experiments utilized the TiZr high pressure cell, allowing pressures up to 3 kbar to be generated using a compressor and capstan pump. Pressure measurement was made using a transducer on the gas pipeline leading to the cell. Data were normalized and corrected using Gudrun to obtain the static structure factor $S(Q)$. The equation of state (EOS) utilized to normalize the data was the Span-Wagner EOS (25) available on NIST REFPROP. Collection time was varied between 8 hours and 24 hours depending on the density of $N_2$ at each pressure studied.

## Empirical potential structure refinement (EPSR)

Neutron diffraction data were analysed using the EPSR software package. At each pressure at which neutron diffraction data were collected a fitting routine was run using a box of 1000 $N_2$ molecules at the fixed density obtained from the Span-Wagner EOS. After allowing the potential energy to equilibrate the empirical potential was turned on and allowed to change in order to fit to the diffraction data. After this process was complete and the energy had stabilized again, the parameters output from EPSR were accumulated for 5000 iterations, keeping the previously obtained empirical potential fixed. Further details on the EPSR analysis method are available in ref. (27). A number of parameters are output by an EPSR simulation. In this work we used the total radial distribution function $g(r)$, the running coordination number and the intramolecular radial distribution function $d(r)$ (simply the distribution of N-N bond lengths). The function $d(r)$ was fitted using a single Gaussian at each pressure studied.

## Ab-initio molecular dynamics (MD) simulations

A series of Ab-Initio molecular dynamics (AIMD) simulations were performed using periodic boundary conditions in a cubic cell containing 108 molecules at densities ranging from 10 mol/l to 45 mol/l. Simulations used the CASTEP code (28) using ultrasoft pseudopotentials with an energy cut off of 490 eV, the PBE exchange-correlation functional and the TS semi-empirical dispersion correction (29). The ab-initio molecular dynamics simulations used a timestep of 1.0 fs and Nosé-Hoover thermostats (5 chains) in a constant NVT ensemble. Each simulation was equilibrated for 1 ps (1000 timesteps) then continued for a further 2000-6000 steps. The pressure for each simulation was obtained from the set density using the Span-Wagner EOS (25). Raman frequencies were obtained by calculating the Fourier transform of each individual molecular bondlength versus time followed by averaging of Fourier transforms over all molecules (Welch windowing and zero padding was used to calculate individual Fourier transforms). In the analysis the Raman frequencies produced using AIMD were normalized by subtraction of a small constant (32 $cm^{-1}$) to ensure the lowest density experimental and AIMD datapoints produced the same Raman shift as the underlying DFT framework resulted in a systematic overestimate of frequency.

The time averaged values of the electronic Hamiltonian are tabulated below, along with the corresponding calculated Raman frequency:

| Density (mol/L) | Frequency ($cm^{-1}$) | Time-averaged Electronic Hamiltonian (Hartrees) |
|---|---|---|
| 10 | 2369.3 | -2.192681 |
| 15 | 2369.7 | -2.192695 |
| 20 | 2365.8 | -2.192740 |
| 25 | 2364.7 | -2.192727 |
| 30 | 2367.1 | -2.192565 |
| 35 | 2369.1 | -2.192256 |
| 40 | 2371.5 | -2.191739 |
| 45 | 2372.6 | -2.190979 |

**Plots of various parameters in $S(Q)$ and $g(r)$ as a function of pressure**

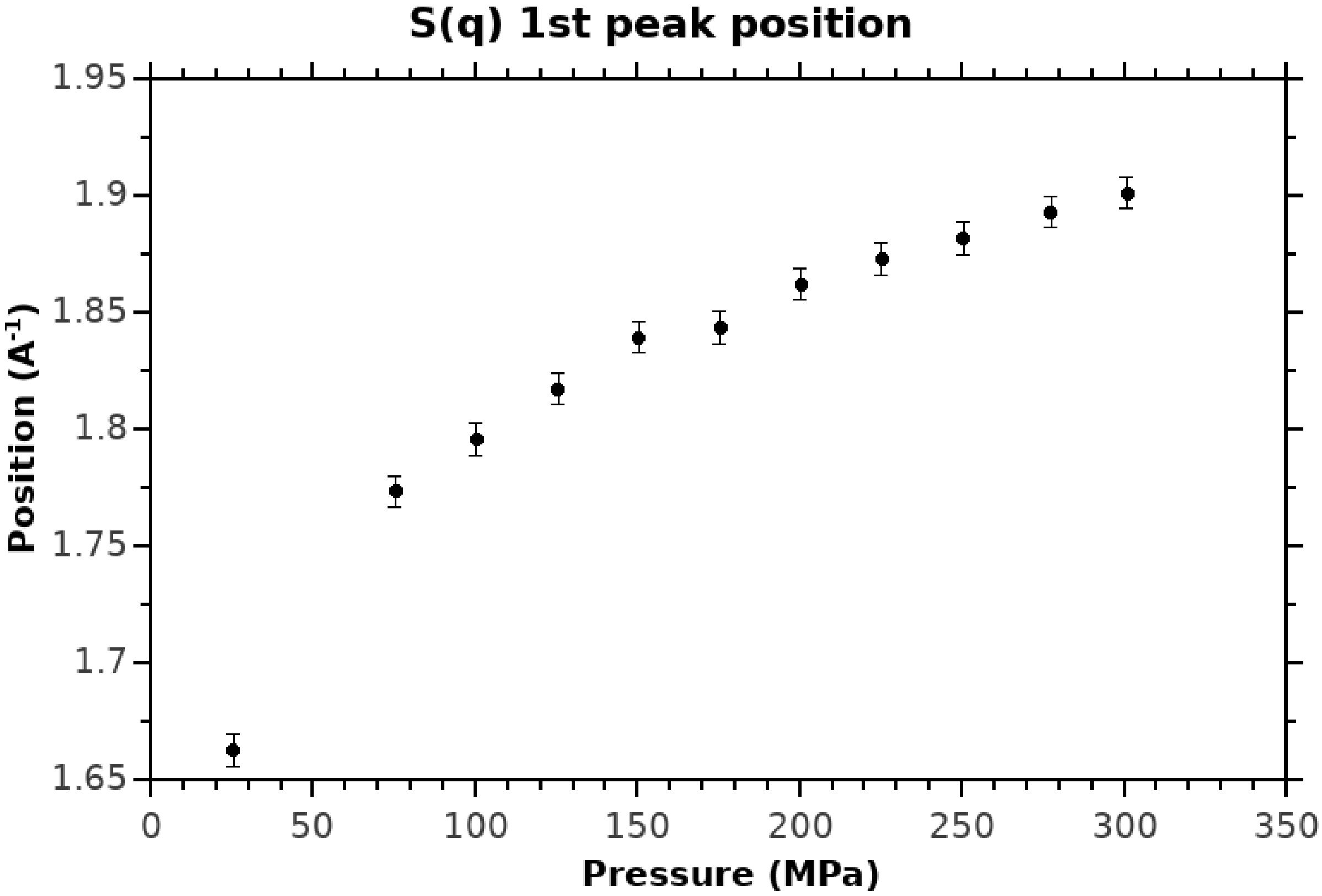

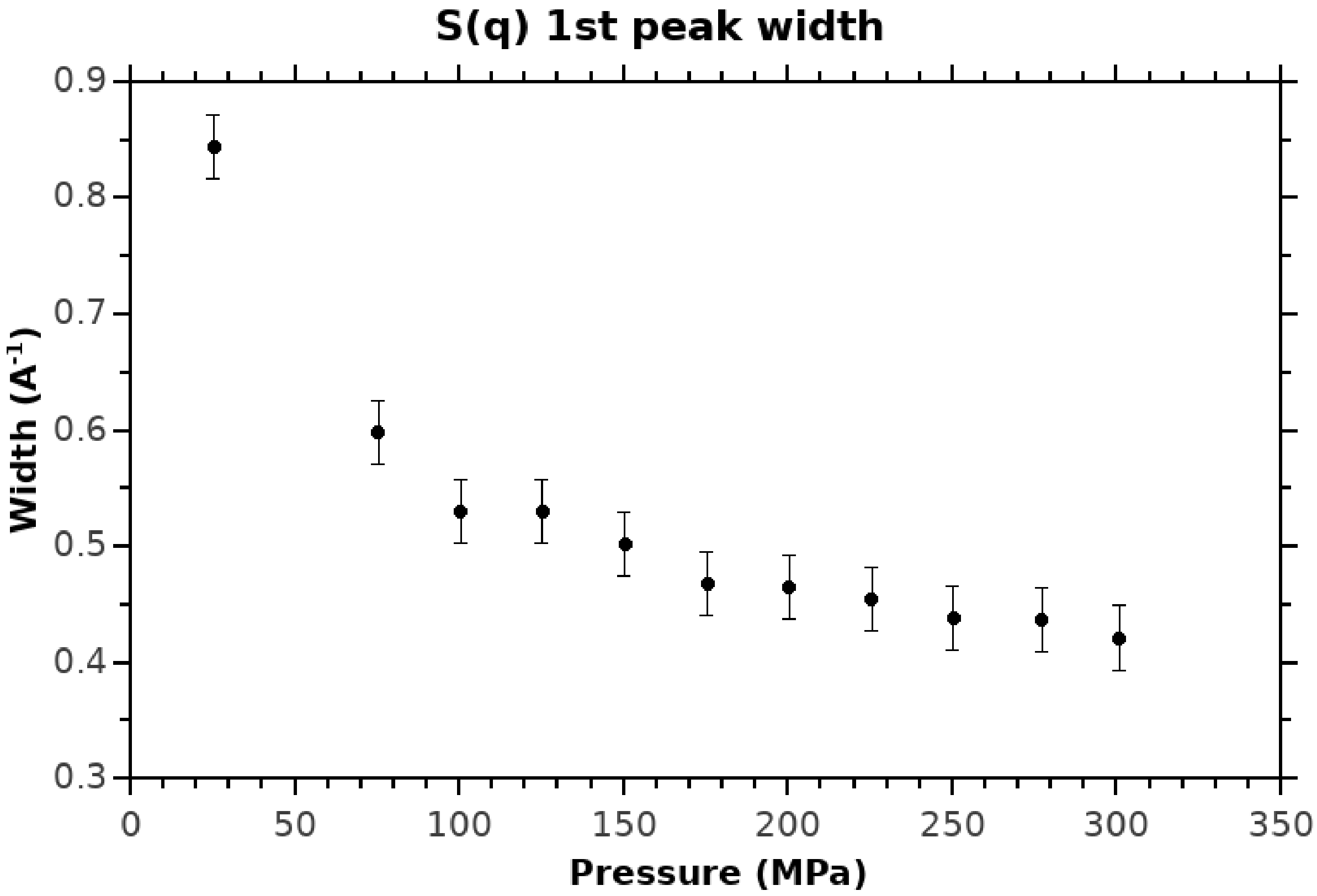

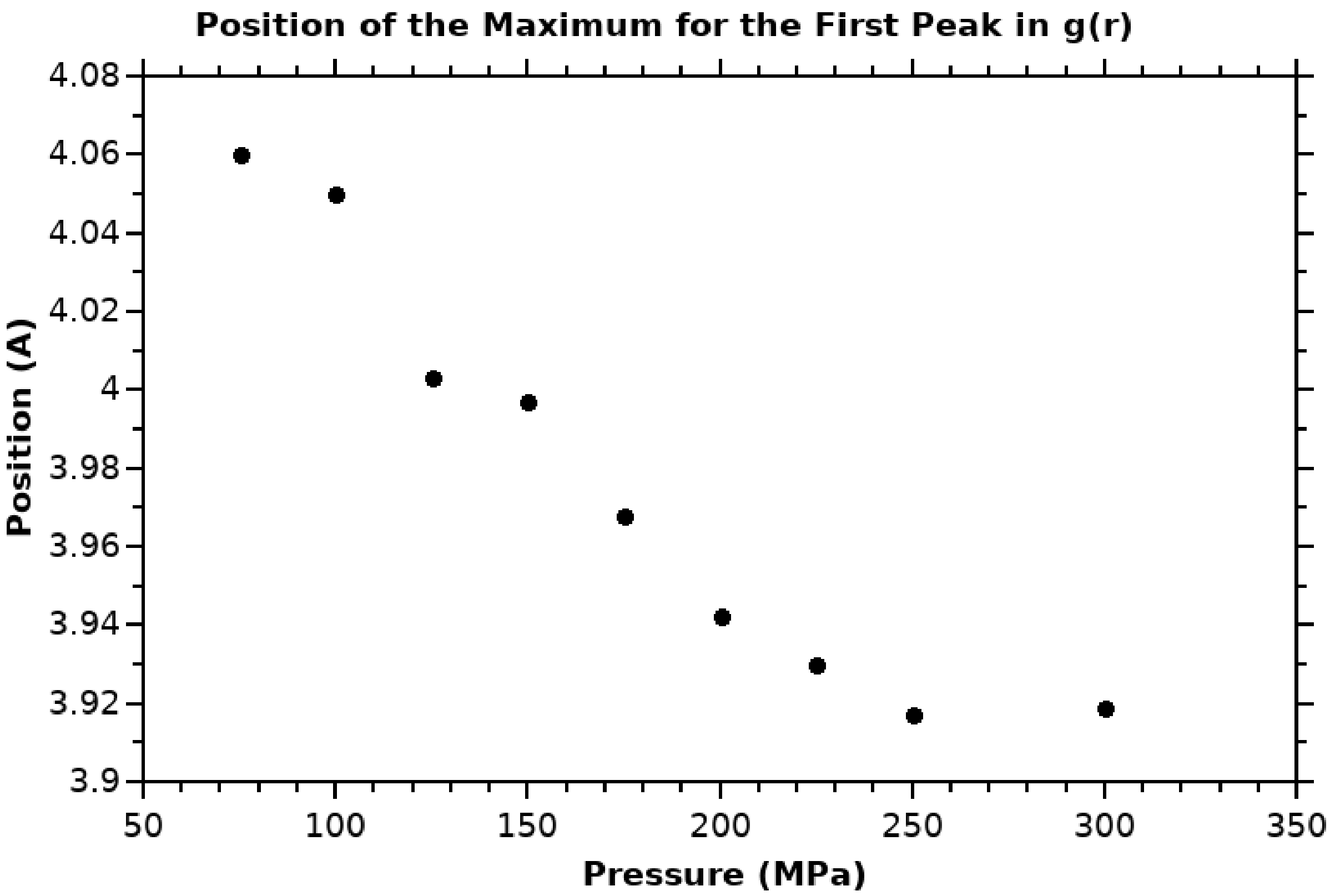
Position of the Maximum for the First Peak in g(r)
Position (A)
4.08
4.06
4.04
4.02
4
3.98
3.96
3.94
3.92
3.9
50
100
150
200
250
300
350
Pressure (MPa)

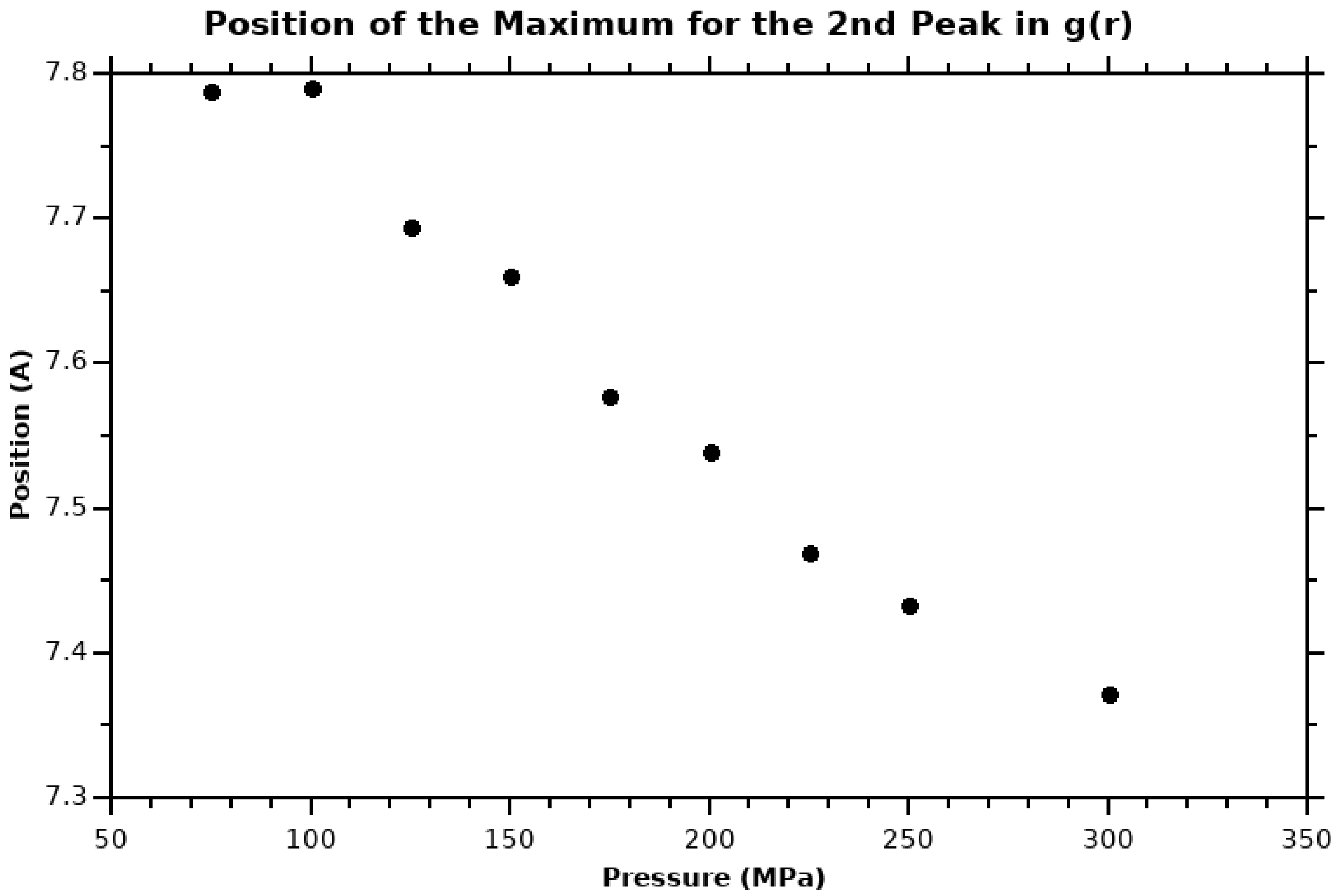
Position of the Maximum for the 2nd Peak in g(r)
Position (A)
7.8
7.7
7.6
7.5
7.4
7.3
50
100
150
200
250
300
350
Pressure (MPa)

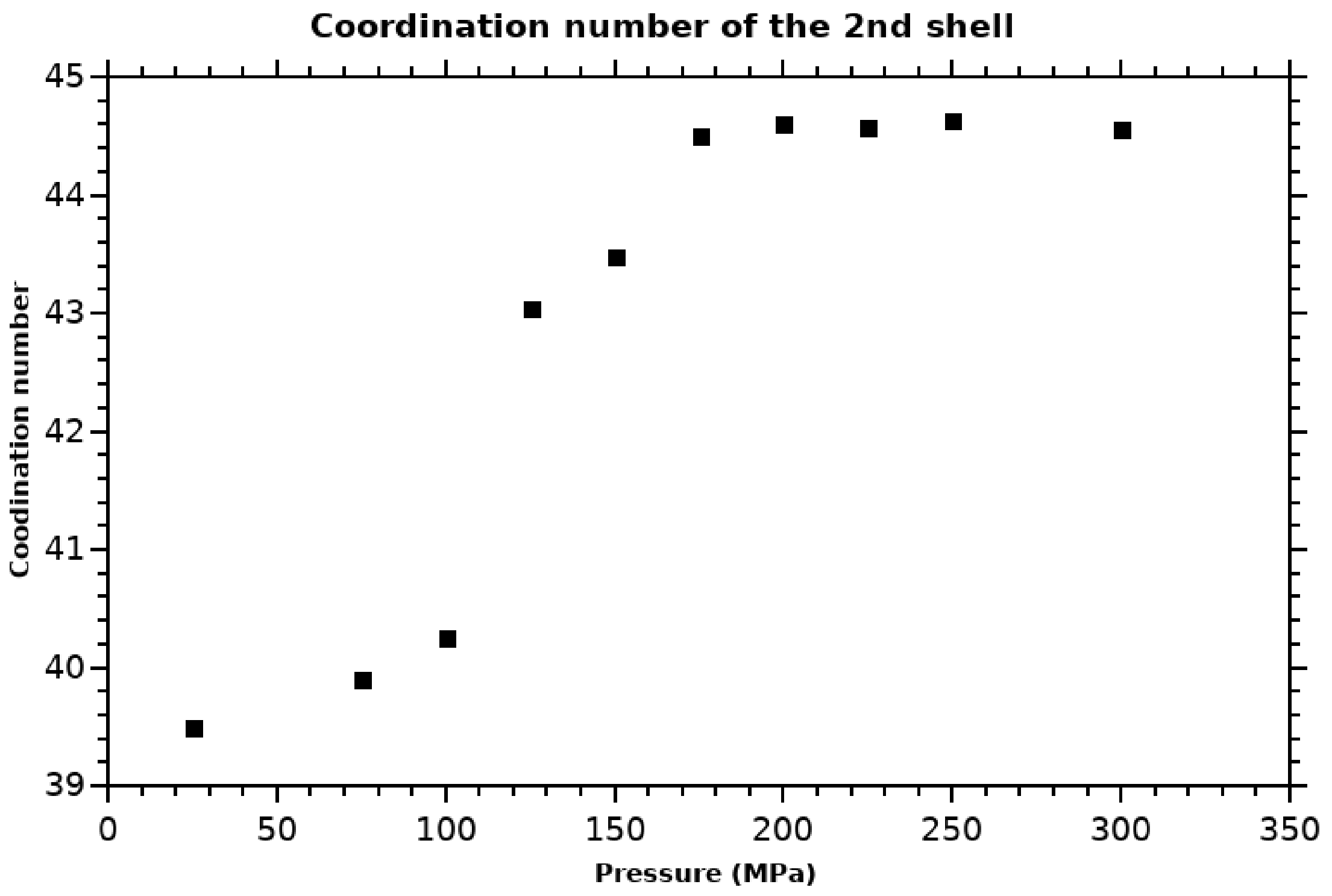

Figure S1. Plots of the position and width of the first peak in $S(Q)$, position of the first and second peaks in $g(r)$ and second shell co-ordination number as a function of pressure.

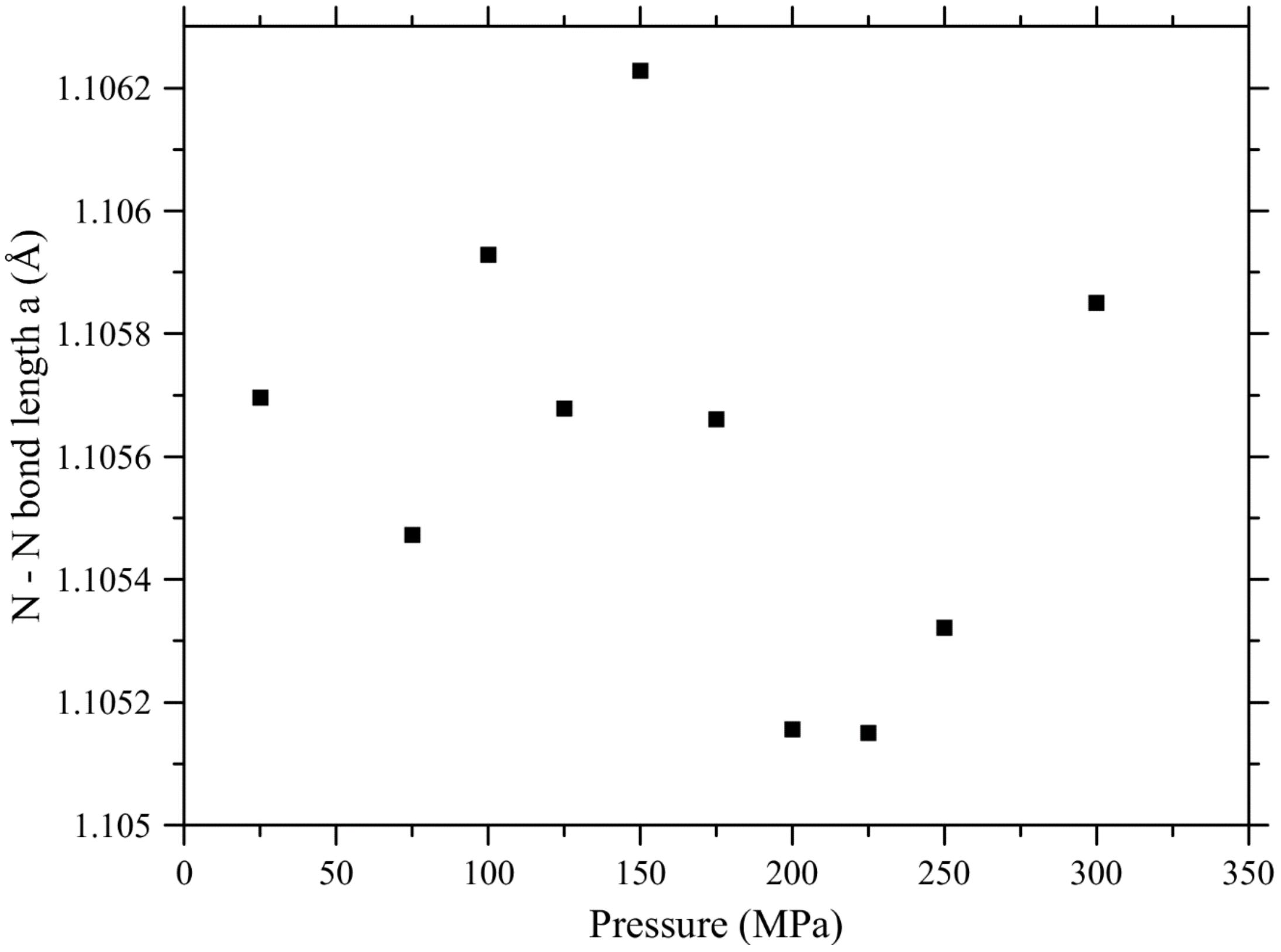

Figure S2. Plot of the N – N bond length obtained from the EPSR software, as a function of pressure.

**Termination of Widom lines**

Out of the Widom lines we studied using the fundamental EOS, those that persisted to the highest temperature were those for viscosity, density (compressibility) and thermal conductivity. In our judgement, by 195 K the variation in these parameters around the critical isochore was not sufficiently large in a narrow $P, T$ range to justify plotting as part of a Widom line. Figure S3 shows plots of these parameters at 195 K.

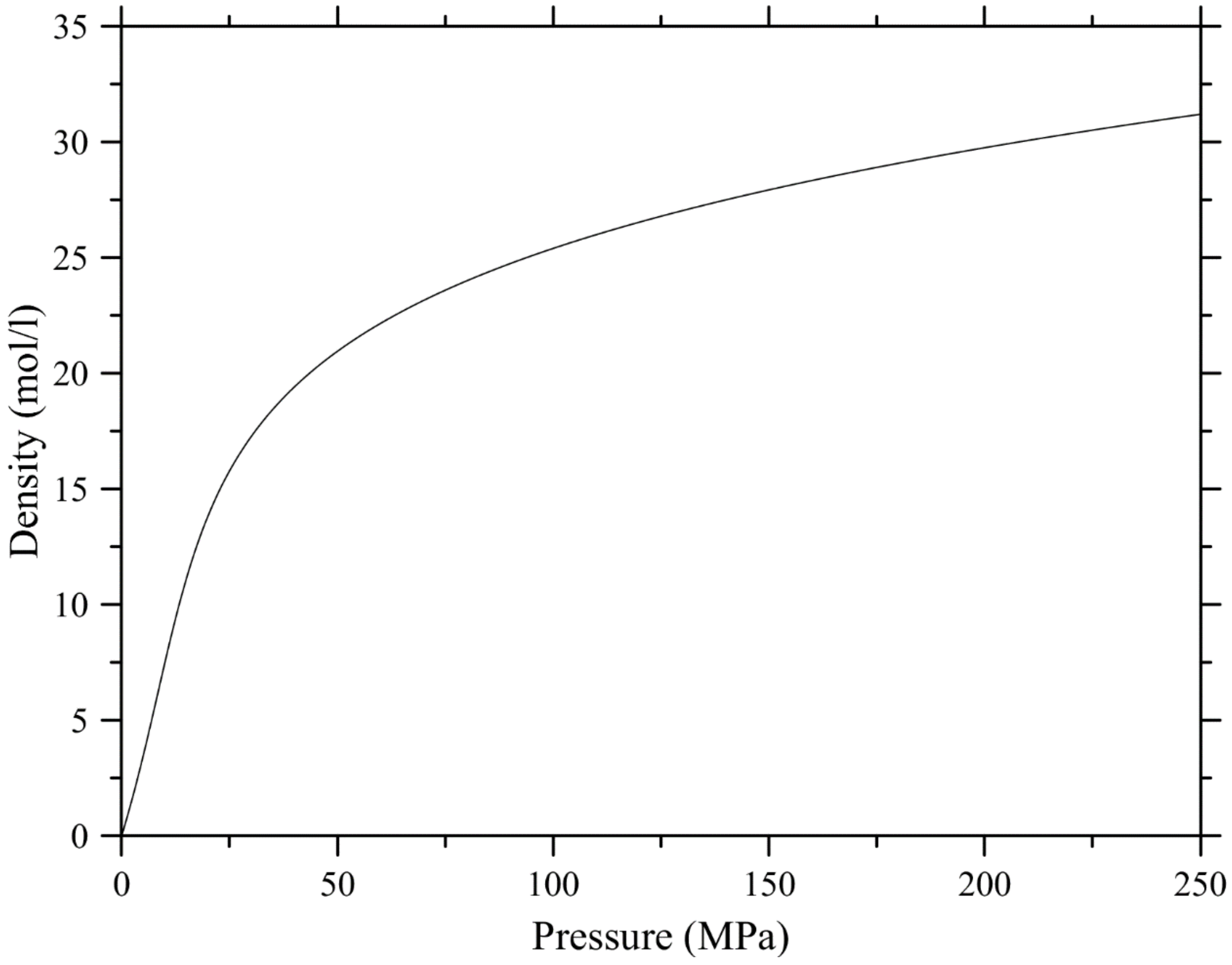

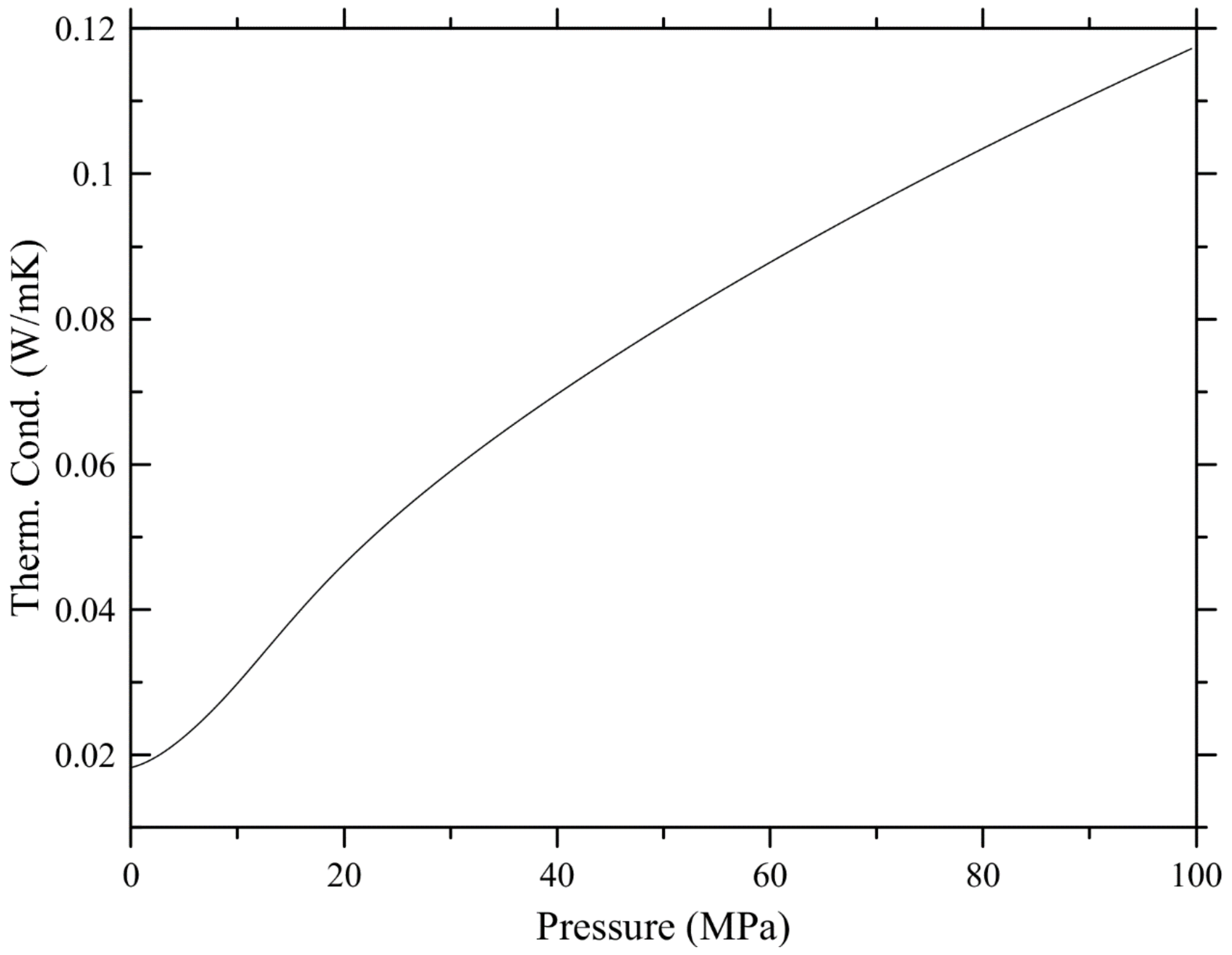

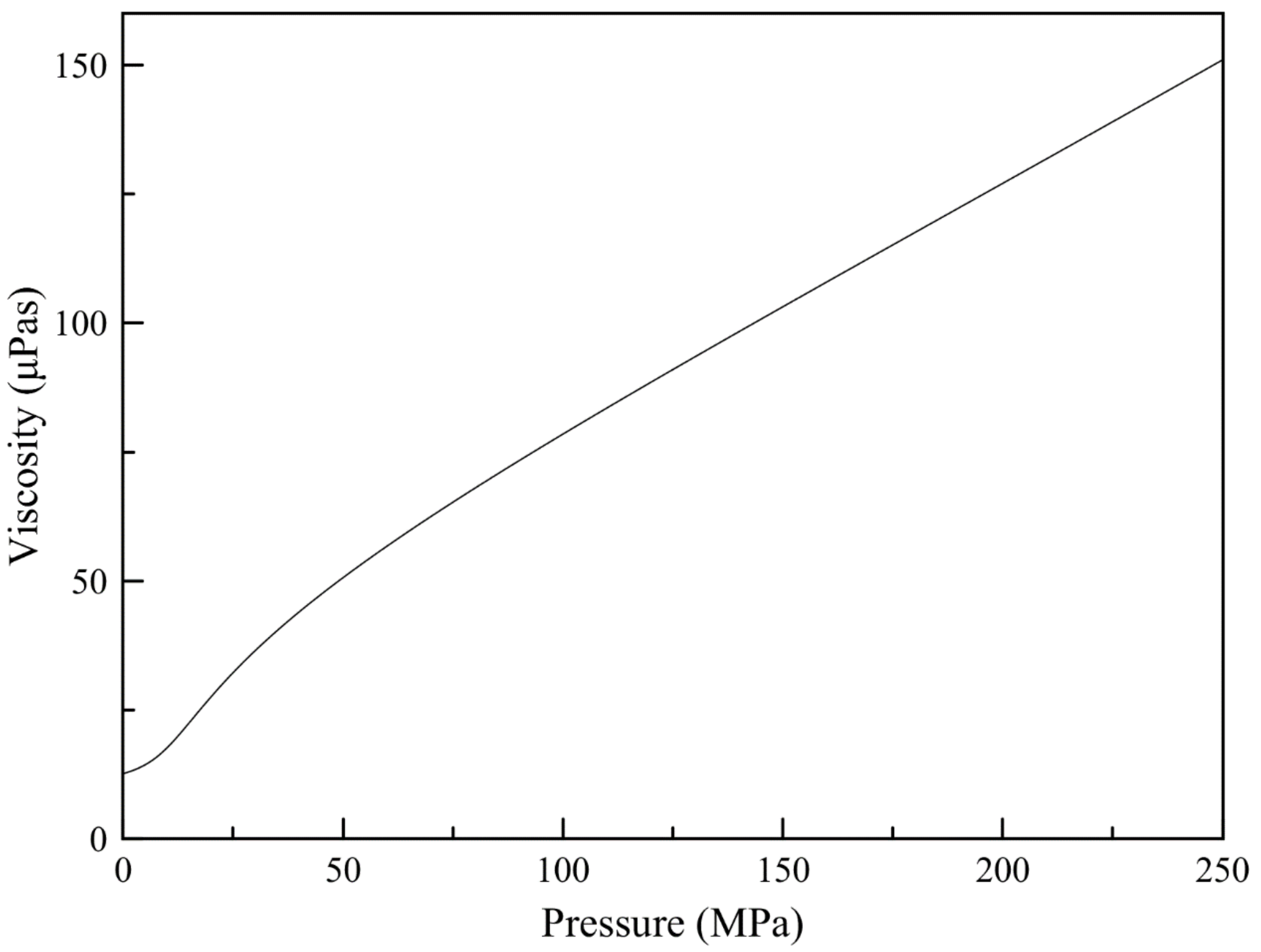

Figure S3. Plots of density, thermal conductivity and viscosity as a function of pressure at 195 K obtained from the fundamental EOS.